# Intrinsic Josephson junction characteristics of $Nd_{2-x}Ce_xCuO_4$/$SrTiO_3$ epitaxial films

T. B. Charikova[1], D.I. Devyaterikov[1], V.N. Neverov[1], M. R. Popov[1*], N.G. Shelushinina[1], A. A. Ivanov[2]

[1]M.N. Mikheev Institute of Metal Physics, Ural Branch, Russian Academy of Sciences, 18, S. Kovalevskoy St., Ekaterinburg, 620108, Russia

[2]National Research Nuclear University MEPhI, Moscow, 115409, Russia

*e-mail: popov_mr@imp.uran.ru

The current-voltage (*I-V*) properties along the *c* axis on $Nd_{2-x}Ce_xCuO_4$/$SrTiO_3$ epitaxial films with $x$ = 0.145, 0.15 were investigated. For all the samples it has been established that the *I-V* characteristics exhibit several resistive branches, which correspond to the resistive states of individual Josephson junctions. The results confirm the idea of a tunneling mechanism between the $CuO_2$ layers (superconductor - insulator - superconductor junction) for the investigated $Nd_{2-x}Ce_xCuO_4$ compound. The *I-V* dependence of this compound with $x$ = 0.15 points out on the nonmonotonic nature of the d-wave or anisotropic *s*-wave symmetry order parameter associated with the coexistence of superconductivity and antiferromagnetic fluctuations.

**Introduction**

Oxide superconductors are layered compounds with building blocks consisting of conductive $CuO_2$ layers separated by buffer layers that serve as charge reservoirs (see monographs [1], [2], [3], [4] for a detailed description). Highly anisotropic high-temperature superconductors (HTSCs) can be considered as a "package" of superconducting $CuO_2$ layers coupled by Josephson interactions [5], [6], [7], [8]. The new properties of these materials compared to single Josephson junctions are associated with their multilayer structure and the atomic thickness of the superconducting layers.

The importance of the mechanism of interlayer coupling (in the *c*- direction) was especially emphasized already at the dawn of the study of cuprate HTSCs, when Anderson developed the model of interlayer tunneling [9]. This model considers tunneling processes along the *c* axis both in the superconducting (tunneling of Cooper pairs) and in the normal (single-particle tunneling) states of a layered superconductor.

Since the transfer between layers occurs through successive tunneling of charge carriers, the corresponding multilayer structures can be considered as a set of intrinsic Josephson tunneling junctions. These junctions are called the intrinsic Josephson junctions [10], [11].

The intrinsic Josephson effect, conditioned by the tunneling of charge carriers in both the superconducting and the normal states of cuprate multilayer HTSCs, has been intensively studied in recent decades. The distinctive structure of the current-voltage characteristics for current in the *c*- direction, with a large (up to several hundred) number of hysteresis resistive branches, is most pronounced in the highly anisotropic Bi- and Tl-systems [12], [13] (see also reviews [10], [11] and references there).

The fact that internal tunneling of Cooper pairs does take place has been experimentally confirmed for many cuprate systems: for $Bi_2Sr_2CaCu_2O_8$ (BSCCO) [10], [11], [12], [13], [14], [15], $Tl_2Ba_2Ca_2Cu_3O_{10}$ (TBCCO) [12], $Bi_2(Sr_{2-x}La_x)CuO_6$ (BSLCO) [16], $La_{2-x}Sr_xCuO_4$ (LSCO) [17], $La_{1.6-x}Nd_{0.4}Sr_xCuO_4$ (LNSCO) [18], for electron-doped cuprates $Pr_{2-x}Ce_xCuO_4$ (PCCO)[19] and $Sm_{2-x}Ce_xCuO_4$ (SCCO) [20], as well as for the magnetic superconductor $RuSr_2GdCu_2O_8$ (RSGCO) [21].

The aim of our work is to investigate intrinsic Josephson effect for the electron-doped cuprate $Nd_{2-x}Ce_xCuO_4$ (NCCO) by measuring the current-voltage characteristics on synthesized $Nd_{2-x}Ce_xCuO_4$ /$SrTiO_3$ epitaxial films.

**Theoretical considerations**

The first, stationary (dc) Josephson effect [22] is as follows: through a tunnel contact between two superconductors (a contact of superconductors through a thin layer of insulator, SIS junction), in the absence of an externally applied potential difference, an undamped (nondissipative) superconducting current (supercurrent) $I_s$ can flow,

$$I_s = I_c sin\Delta\varphi \qquad (1)$$

Here $I_c$ is the critical current of Josephson junction (the maximum value of supercurrent that a Josephson junction can support), $\Delta\varphi$ is the phase difference of the order parameter (of the Ginsburg–Landau wave function) in the two sides of the junction (with the normal state resistance $R$):

$$I_c = \frac{\pi\Delta_0^2}{4eT_cR}, \qquad (2)$$

where $\Delta_0$ is the superconducting gap at $T = 0$, $T_c$ is the critical temperature.

In this case, we are talking specifically about the properties of the junction, and the current $I_c$ is much (by orders of magnitude) less than the destruction current of Cooper pairs in a bulk superconductor (Ginsburg-Landau critical current $I_c^{GL}$): $I_c \ll I_c^{GL}$.

The term *Josephson effect* currently refers to a set of phenomena that occur in contacts of two superconductors through a weak links (see, for example, the review by Likharev [23], as well as the books by Tinkham [3] or Schmidt [24]).

As long as the external current $I$ is less than the critical $I_c$, the entire current is superconducting, $I = I_s$. If $I > I_c$, then in addition to the superconducting current, a normal current component, $I_n$, associated with dissipation, and a voltage $V$ at the contact, appear.

Within the resistive model (see, for example, [24]) for a SIS junction, the Josephson junction can be represented as a parallel connection of an ideal (non-dissipative) contact and the normal resistance $R$. In the general case, it is necessary also to take into account the capacitive effects, since the very design of the Josephson junction, in which the two sides are separated by a dielectric layer, resembles a capacitor. Although direct current is not possible through a capacitor, an alternating bias current $I_d = CdV/dt$ (with $C$ being a capacitance) can flow through it due to recharging of the capacitor plates. A complete description is usually provided by the model of resistively and capacitively shunted junction (RCSJ) [3], [24] (see Fig. 1a).

The total current $I$ through the system is now the sum of the superconducting current $I_s = I_c \sin \Delta \varphi$, the normal dissipative current $I_n = V/R$, where $R$ is the normal state resistance, and the bias current through the capacitor $I_d$:

$$I = I_s + I_n + I_d. \qquad (3)$$

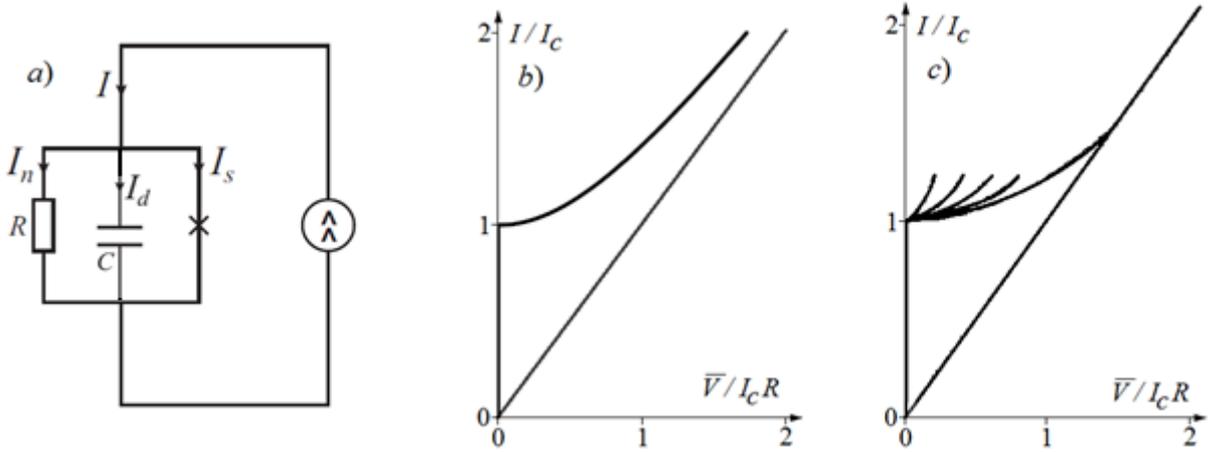

Fig. 1 (a) - RCSJ model of a Josephson junction connected to a circuit with a current source. The Josephson junction is indicated by a cross (from [24]). (b) - Current-voltage characteristic of a single Josephson junction. The voltage at the contact occurs at $I > I_c$ (from [24]). (c)- Generalization to the case of the intrinsic Josephson effect in multilayer system (scheme).

The presence of capacitance in the Josephson contact leads to an ambiguous form of its current-voltage characteristic: the dependence $V(I)$ becomes different for the cases of increasing and decreasing $I$, i.e. hysteresis occurs. Quantitatively, the hysteresis properties are determined by the value of the McCumber parameter [25], [26]:

$$\beta_c = (2e/\hbar)I_c CR^2, \qquad (4)$$

which is a measure of the damping of the Josephson junction.

All of the above applies to a single Josephson junction. The $I$–$V$ characteristics of arrays of Josephson tunnel junctions in multilayer HTSCs consist of several branches, each corresponding to one, two, three, etc individual junctions switching to quasiparticle state as the external current exceeds the corresponding critical current (see reviews [10], [11] and references therein). This situation is shown schematically in Fig. 1c.

The characteristic structure of the current-voltage characteristics for current in the direction of the $c$ axis, with a large number of hysteresis resistive branches is clearly manifested in the most anisotropic layered Bi- and Tl-systems [12], [13].
The dynamic behavior of layered structures with Josephson connections has been analyzed in detail, for example, in [14] and [27] based on a model with multiple sequential Josephson junctions. It is shown that the conditions under which multiple resistive branches are observed on the current-voltage characteristic are associated both with the independent dynamic behavior of each junction and with the presence of hysteresis (with the value of McCumber parameter $\beta$).

**Materials and method**

We were interested in finding out the features of the manifestation of nonlinear dynamic properties of the intrinsic Josephson effect in the electron-doped high-temperature superconductor $Nd_{2-x}Ce_xCuO_4$ (NCCO) with an optimal annealing. The $Nd_{2-x}Ce_xCuO_4$ compound has only one $CuO_2$ plane per unit cell and there are no apex oxygen atoms between adjacent conducting planes in the compounds with optimal annealing (see, for example, [1], [28]). The coupling between cuprate planes is rather weak and a large anisotropy of the conductivity is observed indicating the quasi-two-dimensional nature of the electronic properties.

In our group, $Nd_{2-x}Ce_xCuO_{4+\delta}$/$SrTiO_3$ epitaxial films with $x = 0.145$ ($T_c = 15.7$ K) and 0.15 ($T_c = 23.5$ K) were synthesized by pulsed laser deposition [29], [30]; the $c$-axis of the $Nd_{2-x}Ce_xCuO_{4+\delta}$ lattice in these films is directed perpendicular to the $SrTiO_3$ substrate (standard orientation (001)) and along the long side of the $SrTiO_3$ substrate (orientation ($1\bar{1}0$)).

X-ray diffraction analysis showed that all films with orientation ($1\bar{1}0$) were of predominantly epitaxial. The X-ray diffraction studies of our films were carried out on the PANalytical Empyrean Series 2 diffractometer by the Cu $K_\alpha$ radiation ($\lambda = 1.54$ Å) - measured in parallel beam acquisition geometry. For example, on the Fig. 2 the XRD spectra for $Nd_{2-x}Ce_xCuO_{4+\delta}$/$SrTiO_3$ epitaxial film with $x = 0.145$ is presented. For the film with x = 0.15 were obtained the similar XRD spectra.

Lattice parameters of $Nd_{2-x}Ce_xCuO_4$ apparently are very close to the ones in the substrate, which leads to $(1\bar{1}0)$, $(2\bar{2}0)$ и $(3\bar{3}0)$ plane family Bragg peaks being superposition of two Bragg reflexes from film (red line on the Fig.2) and substrate (blue line on the Fig.2). The signals from the (103) plane family of $Nd_{2-x}Ce_xCuO_4$ presented on diffraction pattern correspond contribution of the c-axis both in-plane and out-of-plane. The detection of interference maxima corresponding to the family of planes (103) and weak interference peaks in the spectrum indicates the presence of structural defects causing angular misorientation ($\leq 1°$) of bulk fragments of the structure. Apparently, this is due to the layered structure of the compound. The obtained diffraction results indicate that the *c*-axis of the compound $Nd_{2-x}Ce_xCuO_4$ predominantly directed parallel to the substrate plane.

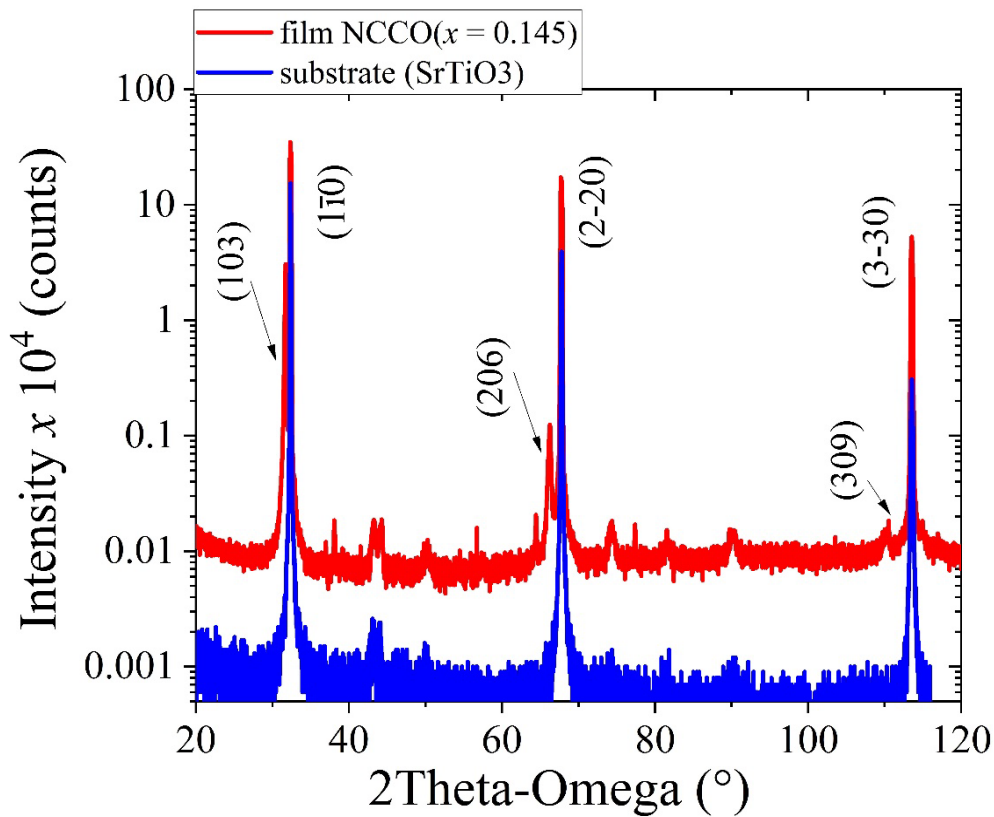

Fig.2. XRD spectra for $Nd_{2-x}Ce_xCuO_4$/$SrTiO_3$ ($x$ = 0.145) epitaxial films with c-axis along the long side of the $SrTiO_3$ substrate (red line) and XRD spectra for $SrTiO_3$ substrate (blue line).

We have used the standard four-probe method; the geometry of the samples was six-contact Hall bar (Fig.3).

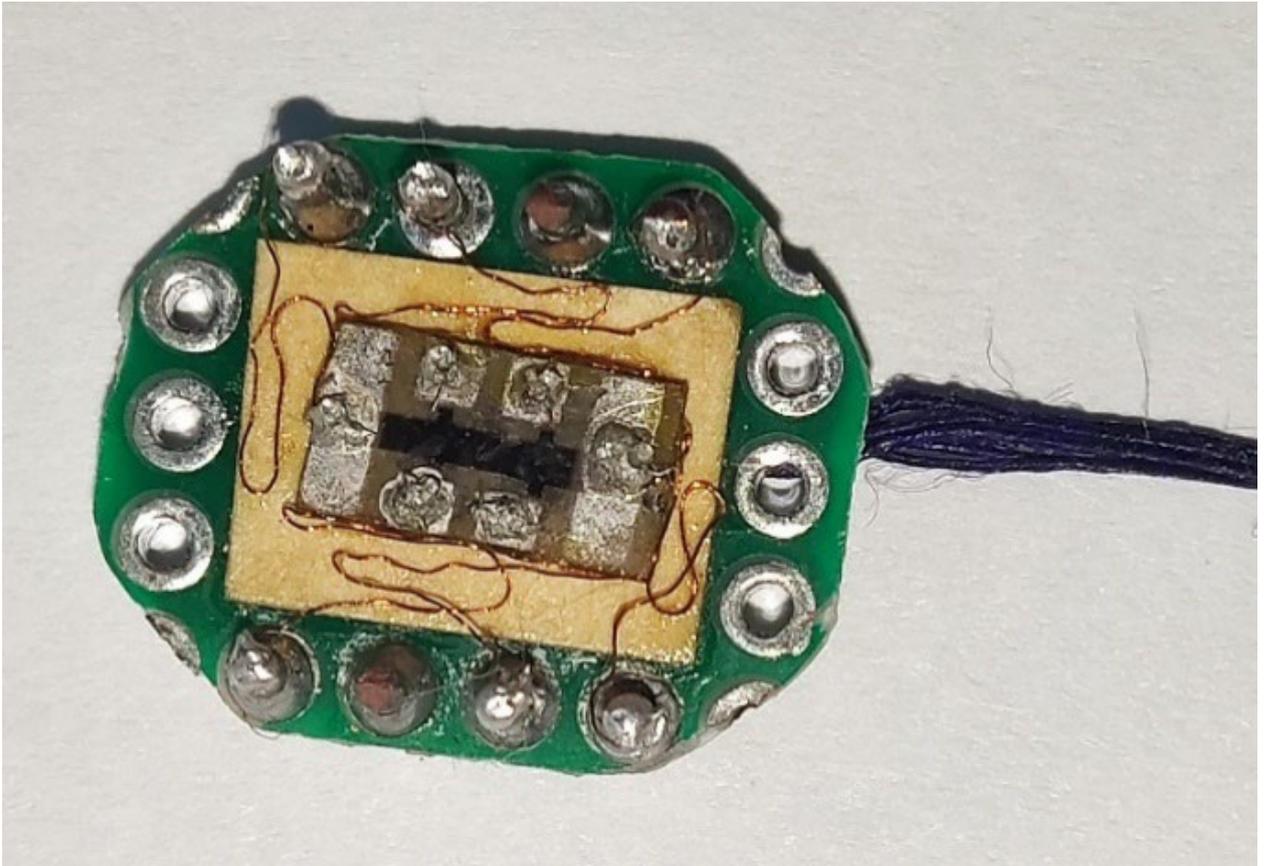

Fig.3. Photo of a sample with contacts installed on the original platform for measuring of the *I – V* characteristics.

The *I–V* characteristics and temperature dependences of the resistivity for all of $Nd_{2-x}Ce_xCuO_4$/$SrTiO_3$ films were investigated in the Quantum Design PPMS 9 and in the original certified setup for measuring of galvanomagnetic effects with the solenoid "Oxford Instruments" (Center for Nanotechnologies and Advanced Materials, IFM UrB RAS). The electric field was applied parallel to the $SrTiO_3$ substrate plane.

**Results and discussion**
*2D metallic state at optimal Ce doping*

In our previous works [31], [32], we have found that in the concentration range $x = 0.145$ and 0.15 $Nd_{2-x}Ce_xCuO_4$ compound is in state of a 2D metal with metallic in nature conductivity in the $CuO_2$ layers and non-metallic conductivity across the layers (Fig.4a,b). The resistance across the $CuO_2$ layers, $\rho_c$, is significantly higher than the resistance, $\rho_{ab}$, in the conductive $CuO_2$ layers and $\rho_c(T)$ reveals a non-metallic temperature dependence for all films studied. It indicates the quasi-two-dimensional character of conductivity in our samples. In this case, the resistivity anisotropy coefficient reaches a value of more than $10^3$ (Fig.4c).

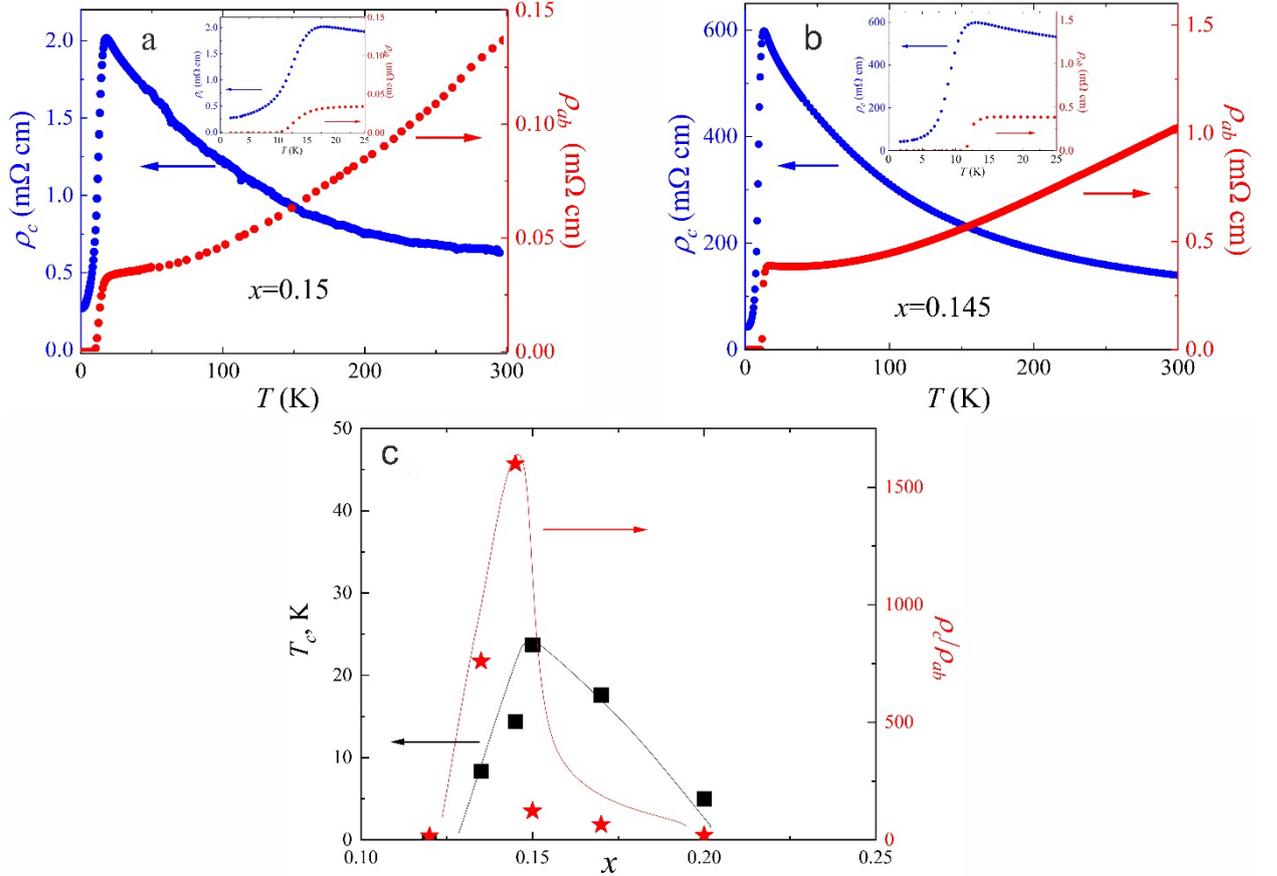

Fig.4. Temperature dependences of in-plane ($\rho_{ab}$) and out-of-plane ($\rho_c$) resistivities for (a) - optimally ($x$ = 0.150) and (b) - almost optimally ($x$ = 0.145) doped Nd$_{2-x}$Ce$_x$CuO$_4$/SrTiO$_3$ films; (c) - dependencies of the critical temperature and the resistivity anisotropy coefficient on dopant Ce content ($x$).

Table 1. Parameters of the Nd$_{2-x}$Ce$_x$CuO$_4$/SrTiO$_3$ epitaxial films. $T_c^{onset}$ is the temperature of the onset of the SC transition, $T_c$ is the SC transition temperature, the ratio $\rho_c/\rho_{ab}$ is given at $T = T_c^{onset}$, $B_{c2}$ is the upper critical magnetic field, $\xi_{ab}$ is the correlation length in the CuO$_2$ plane and $\xi_c$ is the correlation length along the c axis.

| $x$ | $T_c^{onset}$ (K) | $T_c$ (K) | $\rho_c/\rho_{ab}$ | $B_{c2}$ (T) | $\xi_{ab}$ (Å) | $\xi_c$ (Å) |
|---|---|---|---|---|---|---|
| 0.15 | 21.5 | 11.7 | 123 | 6.3 | 72.4 | 6.5 |
| 0.145 | 15.7 | 11.2 | 1600 | 2.7 | 110.0 | 2.9 |

The temperatures $T_c^{onset}$ and $T_c$ are estimated from the $\rho_{ab}(T)$ dependences. Based on the experimental data obtained by us at different times, we were able to estimate the values of the correlation lengths both in the CuO2 plane, $\xi_{ab}$, and along the c axis, $\xi_c$ (see Table 1).

We consider the $Nd_{2-x}Ce_xCuO_4$ crystal as a system of multiple quantum wells ($CuO_2$ layers) separated by doped NdO layers. The situation is similar to that which arises in multilayer heterostructures with selective barrier doping. We have found that two complementary processes determine the *c*-axis transport: incoherent tunneling and thermal activation through barriers [31], [32].

The strong anisotropy of properties in the $Nd_{2-x}Ce_xCuO_4$ layered superconductor, with weak coupling between $CuO_2$ conductive layers, makes it possible to suggest that the $Nd_{2-x}Ce_xCuO_4$ compound is a system with intrinsic Josephson junctions. This statement was proved by us as a result of the study of current-voltage characteristics of $Nd_{2-x}Ce_xCuO_4$ /$SrTiO_3$ epitaxial films.

### *Current-voltage characteristics*

*Preface on intrinsic Josephson junction characteristics in layered cuprates.*
Experimental works on the effects of carrier tunneling along the *c* axis in highly anisotropic HTSCs clearly showed that the materials behave as a set of superconductor-insulator-superconductor (SIS) Josephson junctions (see references in the Introduction). Adjacent superconducting layers in a high-temperature superconductor are weakly coupled by the Josephon effect, causing the single crystals to act essentially as the vertical stacks of hundreds of Josephon junctions.

Typical current-voltage characteristics (*I-V* characteristics) for current flow in the direction of the *c* axis of layered HTSC systems (see reviews [10], [11]) demonstrate a series of sequential Josephson junctions: they have many resistive branches and a clearly defined hysteresis. Current-voltage characteristics of a set of Josephson tunnel junctions consist of several branches, each of which corresponds to one, two, three, etc., individual contacts, which turn into a quasiparticle (normal) state as the applied external current *I* exceeds the corresponding critical current $I_c$.

Almost all the intrinsic Josephson junction (IJJ) characteristics have been observed in hole-doped cuprates and there are only isolated studies for electron-doped cuprates [19], [20]. In an earlier study [19], Schlenga *et al.* observed the *I-V* characteristics of the electron-doped high-$T_c$ superconductor $Pr_{2-x}Ce_xCuO_{4+\delta}$ (PCCO) on a small platelet crystal. They observed the ac Josephson effect and several voltage steps in the *I-V* characteristics, but not a multiple branch structure typical to IJJs.

The paper of Kawakami and Suzuki [20] reports the direct observation of the IJJ characteristics and their multiple branch structure for the electron-doped cuprate $Sm_{2-x}Ce_xCuO_{4-\delta}$ (SCCO) by using a small mesa structure fabricated on a single crystal surface. This result indicates that the crystal structure of SCCO functions as superconductor-insulator-superconductor (SIS-type) IJJs like that of BSCCO. It was found that the *I-V* characteristics in SCCO exhibit multiple resistive branches only when the lateral mesa width is less than a few micrometers.

In this work we investigated the current-voltage characteristics of $Nd_{2-x}Ce_xCuO_4$/$SrTiO_3$ epitaxial films with the *c*-axis of the NdCeCuO lattice (orientation ($1\bar{1}0$)), directed along the long side of the $SrTiO_3$ substrate for two Ce content (*x* = 0.15 and 0.145) at low temperatures *T* = (1.8 – 4.2) K in zero magnetic fields. The observed *I-V* dependences correspond to the standard superconductor-insulator–superconductor tunnel transitions and characterize the $Nd_{2-x}Ce_xCuO_4$ compound as a system with intrinsic Josephson junctions.

*Experimental results for x = 0.15.*

The Fig.5 illustrates the typical *I-V* dependence for optimally annealed $Nd_{2-x}Ce_xCuO_4$/$SrTiO_3$ epitaxial films with *x* = 0.15. The common *I-V* characteristic of optimally annealed $Nd_{2-x}Ce_xCuO_4$/$SrTiO_3$ epitaxial film (*x* = 0.15) at *T* =1.8 K (Fig.5a) and *T* = 4.2 K (inset on Fig.5c) demonstrates a zero-voltage supercurrent at $I < I_c$ with a low critical current value $I_c \cong 10 \mu A$.

In the *I-V* characteristic for the current in the *c* – direction the signs of intrinsic Josephson transitions are displayed in the region of $I > I_c$: N = 3 resistive branches are observed, separated by equal intervals δU =31.5mV at I = 50μA (see the insert on Fig. 5b). The linear dependence $U(N)$ indicates the uniformity of the junctions in the film. It should be noted the absence of hysteresis effects for the studied sample with *x* = 0.15: the *I-V* dependences are reversible for a given branch when the current increases and then decreases to zero.

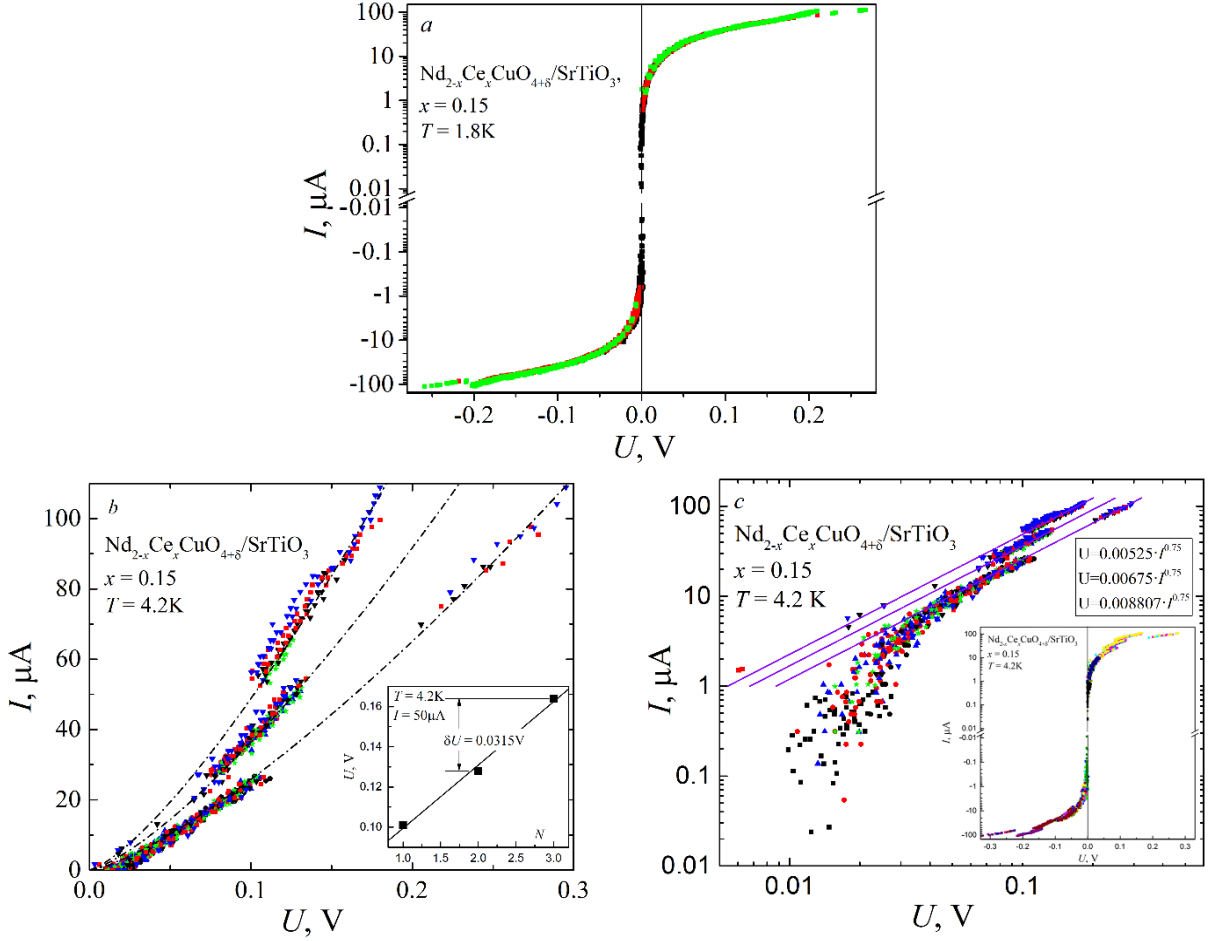

Fig.5. (a) - The common *I-V* characteristic of optimally annealed $Nd_{2-x}Ce_xCuO_4$ /$SrTiO_3$ epitaxial film ($x = 0.15$) at $T = 1.8$ K; (b) - *I-V* characteristic with three resistive branches for intrinsic tunnel junctions at $T = 4.2$ K. Dashed lines correspond to the dependence $U \sim I^{0.75}$. Inset: the voltage measured at 50 µA for every counted branch (N = 1, 2, 3); (c) - a scaling behavior of intrinsic *I-V* characteristics for different branches of $Nd_{2-x}Ce_xCuO_4$ /$SrTiO_3$ epitaxial film ($x = 0.15$) at $T = 4.2$ K. Inset: the common *I-V* characteristic of the film at $T = 4.2$ K.

In Fig. 5c the *I-V* dependence for the film with $x = 0.15$ is presented on a double logarithmic scale. We see that when $I > \sim 5$ µA experimental data for all three branches fit on straight lines corresponding to the dependence $U \sim I^{0.75}$. Dependencies of this type are given in Fig. 5b by the dashed lines. The spread of experimental points at very low currents $I < \sim 5$ µA in Fig. 5c is due to random fluctuations in the values of the critical current $I_c$ when the transition to a resistive state is turned on.

*Discussion of results for x = 0.15.*

Let us discuss the functional dependence of *V* on *I* for each of the branches in the current-voltage characteristic on Fig. 5c. We see that the current rise is gradual for all branches with a high conductivity in the subgap regime. For the case of tunneling between two superconductors with BCS-like density of states (for conventional s-wave symmetry of the order parameter), a steep current rise should take place at a voltage $2\Delta/e$, where $\Delta$ is the superconducting gap [10]. Thus, the resistive branches should be almost vertical at these voltages and the current below the gap (subgap current) should be inessential at low temperatures, $I \sim \exp(-\Delta/kT)$.

The tunneling characteristics of the intrinsic Josephson junctions in the hole-doped BSCCO as well as in TBCCO at low temperatures show a high conductivity in the subgap regime $V < 2\Delta/e$ (see [33] and other references below). A large number of experimental results testifies in favor of an idea about predominantly d-wave pairing symmetry in optimally hole-doped cuprates ($d_{x2-y2}$ - symmetry of the order parameter) [34], [35].

An existence of the nodes in the gap for the d-wave order parameter leads to a considerable change in the I-V tunnel characteristics: a tunnel junction of d-wave superconductors should show a large subgap current in contrast to the s-wave case [36], [37].

Within the confines of standard tunneling formalism, quasiparticle current for SIS tunnel junction with $d_{x2-y2}$-wave density of states for the superconductors in the $CuO_2$ planes is numerically calculated both for coherent [38] and incoherent [33], [38] tunneling cases. The coherency (or incoherency) of tunneling process is defined by the conservation (or nonconservation) of the transverse momentum. As a whole, calculations indicate that for $V < 2\Delta/e$ the subgap conductance $\sigma(V) \equiv dI/dV$ is almost linear in $V$ in the coherent tunneling case [38], while $\sigma(V)$ is nearly proportional to $V^2$ for incoherent tunneling [33], [38]. Then, theoretically, for a general curvature of a branch in the current-voltage characteristic of SIS tunnel junction with a superconducting order parameter that has a predominant $d_{x2-y2}$ symmetry we should have $V \sim I^\gamma$, where $\gamma \approx 1/2$ for coherent tunneling and $\gamma \approx 1/3$ for incoherent case.

In [33] and [38] a numerical fit of the calculated I-V dependences to selected experimental curves was carried out. To reveal the general curvature of I-V dependences for branches in the current-voltage characteristic we presented a large set of relevant experimental data for hole-doped cuprate HTSCs on a double-logarithmic scale, similar to our Fig. 5b for NCCO.

We used data of works [10], [33] for Tl - systems and of works [11], [39] for Bi - systems. It was found that for such a set of hole-doped samples the I-V curves for individual branches in the current-voltage characteristic fits the power-law dependence $V \sim I^\gamma$ with $\gamma = (0.29 - 0.38)$ in good agreement with theoretical calculations ($\gamma \approx 1/3$) for the case of incoherent tunneling. For high-quality $Bi_2(Sr_{2-x}La_x)CuO_6$ mesa-structures [16] the treatment of V-I dependences for an array of resistive branches leads to the value $\gamma = 0.48$, which may correspond to the case of coherent tunneling processes with $\gamma \approx 1/2$ [38].

It is logical to associate the much larger values of $\gamma$ (= 0.75) we observed for NCCO with another type of d-symmetry of the order parameter, characteristic for electron-doped cuprates [40], or, as a possible option, with a case of anisotropic *s*-wave symmetry [41]. For the electron-doped cuprates with optimal doping the interpretation of a great deal of experimental facts is consistent with *nonmonotonic* form of the *d*-wave superconductor order parameter (see [41], [42] and a great number of references in review of Armitage et al. [40]) as opposed to hole-doped cuprates with simple *monotonic* $d_{x2-y2}$ symmetry.

*Experimental results for x = 0.145.*

The *I-V* characteristics for the optimally annealed $Nd_{2-x}Ce_xCuO_4$ /$SrTiO_3$ epitaxial film (*x* = 0.145) at *T* = 1.8 K demonstrates multiple branches in the resistive state at $I > I_c$ (Fig.6a, b, c). The *I-V* curves are hysteretic that typical for Josephson tunnel junctions with large capacitance [24]. The precise *I-V* measurements made it possible to visualize the jump from one resistive branch to other with an increase in current ($I = 0 \to I^*$) and back from one to other with a decrease in it ($I = I^* \to 0$) with $I^* = 20, 30, 45, 60, 70, 80$ and $90$ µA (see the top insert in Fig. 6b).

Although the quality of the resulting *I-V* curves for *x* = 0.145 is for some reason a little worse than for *x* = 0.15, they clearly show the regularly spaced branches (with $\delta U = 0.2 mV$ at 50 µA) and it's possible to determine the numbers of observed junctions: $N = 1, 3, 4, 5$ (see the bottom insert in Fig. 6b). The linear dependence $U(N)$ indicates the uniformity of the junctions in the film. This behavior corresponds to the standard superconductor-insulator–superconductor tunnel junction and characterizes the $Nd_{2-x}Ce_xCuO_4$ compound as an array of resistively and capacitively shunted intrinsic Josephson junctions.

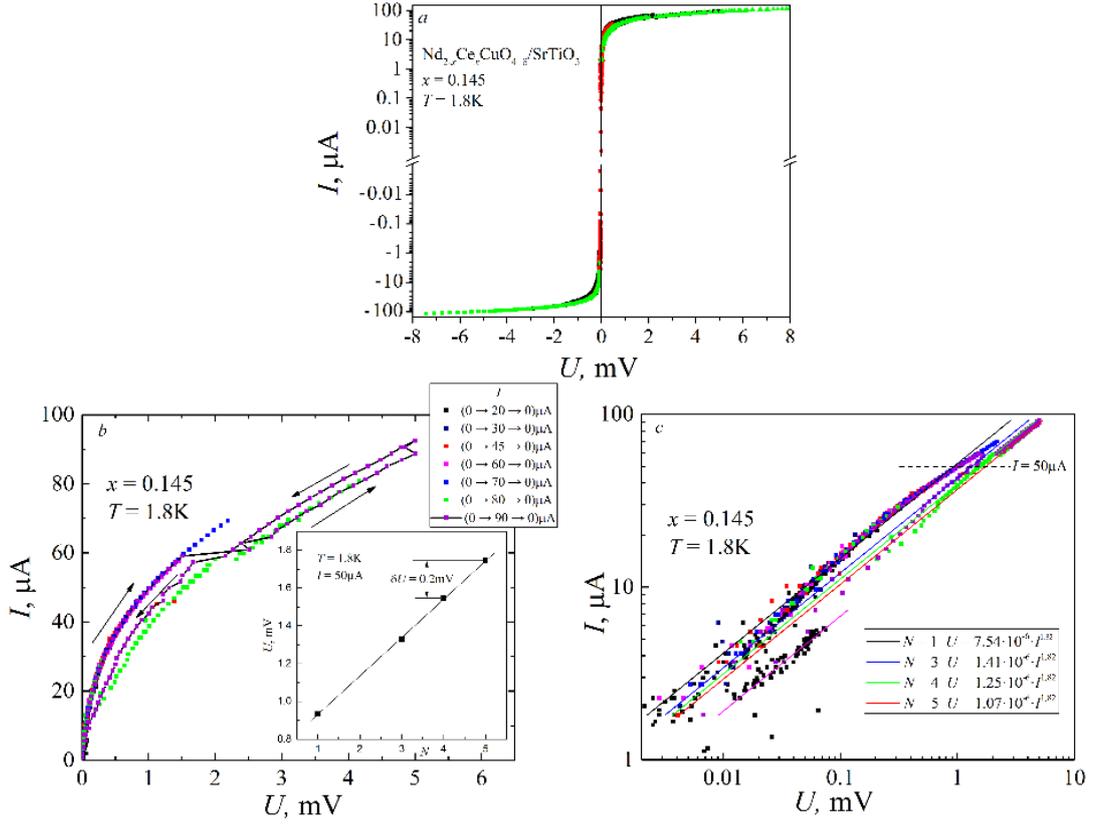

Fig.6. (a) - The common *I-V* characteristic for optimally annealed $Nd_{2-x}Ce_xCuO_4/SrTiO_3$ epitaxial film ($x = 0.145$) at 1.8 K with zero-voltage supercurrent at $I < I_c \cong 40\mu A$.; (b) - hysteretic *I-V* dependence and jumps from one resistive branch to other for this type of films at $T = 1.8$ K. Inset: the voltage measured at 50 μA for every realizable branch (N = 1,3,4,5); (c) - low temperature scaling behavior of intrinsic *I-V* characteristics for optimally annealed $Nd_{2-x}Ce_xCuO_4/SrTiO_3$ epitaxial film ($x = 0.145$) at $T = 1.8$ K. The lowest line based on the totality of black dots corresponds to the resistive branch with N = 15.

*Discussion of results for x = 0.145.*

Several distinctive features should be noted in the behavior of the *I-V* characteristics for the sample with $x = 0.145$:

i) the observation of hysteresis effects in a sample with $x = 0.145$ is usually associated with a significant capacitive contribution to the current in the resistive state, the measure of which is expressed in the value of the McCumber parameter $\beta_c$ [25], [26]. Let's estimate the relationship between these parameters for $x = 0.145$ ($\beta_c^{(1)}$) and $x = 0.15$ ($\beta_c^{(2)}$).

The general formula for the McCumber parameter is (see Eq.(4)):

$$\beta_c = \left(\frac{2e}{\hbar}\right) I_c C R^2 = \frac{\pi}{2\hbar} \frac{\Delta_0^2}{T_c} RC,$$

where we used Eq. (2) for critical current. Then we have:

$$\frac{\beta_c^{(1)}}{\beta_c^{(2)}} = \frac{T_c^{(1)}}{T_c^{(2)}} \frac{\rho_c^{(1)}}{\rho_c^{(2)}} \cong 150, \tag{5}$$

where $\rho_c(Tc)$ values from Figs 4a,b for $x = 0.145$ ($\rho_c^{(1)} = 600 m\Omega \cdot cm$) and $x = 0.15$ ($\rho_c^{(2)} = 2 m\Omega \cdot cm$) were used.

The estimates give $\beta_c^{(2)}$ of the order of unity, then $\beta_c^{(1)} > 100$ and in accordance with numerical calculations of Stewart [26] it leads to a complex form of hysteresis current-voltage characteristic;

ii)    much smaller intervals between individual branches of intrinsic Josephson junctions ($\delta U^{(1)}$=0.2mV at $I$ = 50 µA) in comparison with the intervals for the sample with $x$ = 0.15 ($\delta U^{(2)}$=31.5mV). In principle, this may be due to resonant transitions through an impurity in the barrier [43], but most likely it is a result of the capacitive effects. However, the reason may also lie in the technology of growing the films under study (see results of the X-ray diffraction studies above).

On the one hand, the measured *I-V* characteristics on Figs. 6b, c are mainly due to the conductivity along the *c*- axis (out-of-plane contribution), since we observe a set of *I-V* branches from the intrinsic Josephson junctions. On the other side, a contribution of in-plane conductivity due to a slight misorientation of the *c*- axis relative to the substrate can lead to effective shunting of the tunneling process between the $CuO_2$ planes which may be expressed in a decrease of the $\delta U$ intervals between the *I-V* branches. This effect may be more pronounced for the $x$ = 0.145 film due to a large anisotropy, $\rho_c/\rho_{ab} = 1600$, and may be less important for the $x$ = 0.15 film with a much smaller degree of anisotropy, $\rho_c/\rho_{ab} \cong 100$;

iii)    in the dependences $U \sim I^\gamma$ observed for $x$ = 0.145 the value of $\gamma > 1$ (see Fig.6c, $\gamma \cong 1.8$), that is, the *I-V* curves are convex in contrast to the concave *I-V* curves for $x$=0.15. A significant contribution of capacitive effects in the resistive state of Josephson junctions due to the significantly larger value of the McCumber parameter for $x$ = 0.145 may play determining role and for the form of *I-V* dependence.

Summarizing the results and the assessments made, we represent the parameters of the samples related to the studied volt-ampere characteristics in a Table 2. For the samples of the $Nd_{2-x}Ce_xCuO_4$ with $x$ = 0.15 and 0.145, Table 2 shows the following parameters: the values of the resistivity $\rho_c$ at $T = T_c^{onset}$ (see Fig. 4a, b), of the resistance in the normal state $R = \rho_c \frac{c}{S}$ and of the capacitance $C = \varepsilon \frac{S}{c}$ for a single junction, where $c = 60\ nm$ is the separation between $CuO_2$ layers and $S$ is the cross-sectional area of the sample, $\varepsilon = \varepsilon_0 \cdot \varepsilon_r$, $\varepsilon_0$ is the *electric constant* ($\varepsilon_0 \sim 8.85 \times 10^{-12} F \cdot m^{-1}$) and $\varepsilon_r$ is the dielectric constant of the material (we use for NCCO $\varepsilon_r \cong 10$ [4]). In addition, Table 2 shows the experimentally observed values of the critical current for Josephson junction systems, $I_c$, at 1.8 K (see Fig. 5a and Fig. 6a), as well as the value of the McCumber parameter, $\beta_c$, calculated using formulas (4) and (2), and the values of the exponent, $\gamma$, in the dependences $V \sim I^\gamma$.

Table 2. Parameters of the Nd$_{2-x}$Ce$_x$CuO$_4$ samples.

| $x$ | $\rho_c (m\Omega \cdot cm)$ | $R(10^{-5}\Omega)$ | $C(10^{-11}F)$ | $I_c(\mu A)$ | $\beta_c$ | $\gamma$ |
|---|---|---|---|---|---|---|
| 0.15 | 2 | 4.3 | 4.2 | 10 | ~1 | 0.75 |
| 0.145 | 600 | 1125 | 4.7 | 40 | ~100 | 1.8 |

From the analysis of the data in Tables 1 and 2 we firstly see the approximately twofold difference in the temperatures of SC transition, $T_c$, what is due to the different Ce content in the samples. And, most importantly, there is a great difference (300 times) in the values of the resistivities $\rho_c$ at $T = T_c^{onset}$, which further leads to a difference in the values of $R$ (260 times), and, importantly, in the assessments of McCumber parameter $\beta_c$ (~100 times) in favor of $x = 0.145$ structure. The sharp increase of the normal state resistivity $\rho_c(T)$ when approaching $T_c$, together with a strong resistive anisotropy (see Figs 4b, c) indicates the high quality of the layered structure in the grown Nd$_{2-x}$Ce$_x$CuO$_4$/SrTiO$_3$ film with $x = 0.145$.

It is known [25], [26] that the McCumber parameter is a measure for relative contributions of resistive and capacitive effects to the shunting of the Josephson junction at $I > I_c$ (see Eq. (3)). Strong differences in the value of the McCumber parameter ($\beta_c \sim 1$ for $x = 0.15$ and $\beta_c \sim 100$ for $x = 0.145$) lead to the implementation of different limiting cases of the RCSJ model in the studied samples (see numerical calculations in [26]).

For a case of $\beta_c \sim 1$, at $I > I_c$ the resistive component dominates in Eq.(3) ($I_n > I_r$), but for $\beta_c \sim 100$, on the contrary, the capacitive contribution plays the decisive role in the *I-V* behavior ($I_r \gg I_n$). Thus, the RSJ (resistively shunted junction) case is realized in $x=0.15$ sample and the CSJ (capacitively shunted junction) case is settled in $x = 0.145$ sample, which leads to significantly different functional *I-V* dependences in these samples, in particular, to significantly different values of the parameter $\gamma$.

**Conclusions**

We have found that in low-dimensional electron doped superconducting system Nd$_{2-x}$Ce$_x$CuO$_4$ with large-scale quantum coherence, the properties of superconducting weak links (Josephson junctions) appear on a macroscopic scale. It has been established that the current-voltage characteristics exhibit several resistive branches, which correspond to the resistive states of individual Josephson junctions. The results confirm the idea of a tunneling mechanism between the $CuO_2$ planes (superconductor - insulator - superconductor transitions) for the investigated $Nd_{2-x}Ce_xCuO_4$ compound [44].

This paper presents the studies on $Nd_{2-x}Ce_xCuO_{4+\delta}$/$SrTiO_3$ films ($x = 0.145$ and $0.15$) synthesized by pulsed laser deposition with the *c*-axis of the $Nd_{2-x}Ce_xCuO_{4+\delta}$ lattice directed along the long side of the $SrTiO_3$ substrate (orientation ($1\bar{1}0$)). X-ray diffraction analysis indicate that the *c*- axis of the compound $Nd_{2-x}Ce_xCuO_4$ predominantly directed parallel to the substrate plane with a slight angular misorientation due to a presence of structural defects.

The investigated *I-V* characteristics correspond to the process of the conduction along the *c*- axis (out-of-plane contribution): a set of branches from an array of intrinsic Josephson junctions are observed. Some in-plane contribution due to a slight misorientation of the *c*- axis relative to the substrate can lead to a shunting of the tunneling process between the $CuO_2$ planes which may turn out to be effective for the $x = 0.145$ film with a high degree of anisotropy, $\rho_c/\rho_{ab} \sim 10^3$.

One can consider $Nd_{2-x}Ce_xCuO_4$ single crystal as a system of multiple quantum wells ($CuO_2$ layers) separated by doped NdO layers or just as periodic multibarrier tunneling structure that are, by geometry, reminiscent of 2D semiconducting multilayer heterostructure [45], [46]. An overall transmission coefficient for tunneling through such structures is known to have several sharp peaks including the superconducting properties in these systems for the *s*-symmetry of the SC order parameter. But for the SC pairing symmetry of *d*-type some subgap structure should be evident.

The previous high-quality studies (see the review of Armitage [40]) support the version of nonmonotonic nature of the *d*-wave order parameter in NCCO associated with the coexistence of superconductivity and antiferromagnetic fluctuations in the region of optimal doping. As a result of research of the current-voltage characteristics, carried out by us, the features of the type of pairing symmetry in electron-doped cuprate $Nd_{2-x}Ce_xCuO_4$ with optimal annealing were analyzed.

**Acknowledgments**